\documentclass[conference]{IEEEtran}

\newtheorem{theorem}{Theorem}
\newtheorem{lemma}{Lemma}
\newtheorem{corollary}{Corollary}
\newtheorem{definition}{Definition}
\newtheorem{proposition}{Proposition}
\newtheorem{remark}{Remark}

\usepackage[cmex10]{amsmath}
\usepackage{amssymb}
\usepackage{subfig}

\usepackage[dvips]{graphicx}

\IEEEoverridecommandlockouts

\begin{document}
\title{On a Class of Discrete Memoryless Broadcast Interference Channels}

\author{\IEEEauthorblockN{Yuanpeng Liu, Elza Erkip}
\IEEEauthorblockA{ECE Department, Polytechnic Institute of New York University\\
yliu20@students.poly.edu, elza@poly.edu}
\thanks{This work was supported in part by NSF grant No. 0635177 and InterDigital Communications, LLC.}}

\maketitle

\begin{abstract}
We study a class of discrete memoryless broadcast interference channels (DM-BICs), where one of the broadcast receivers is subject to the interference from a point-to-point transmission. A general achievable rate region $\mathcal{R}$ based on rate splitting, superposition coding and binning at the broadcast transmitter and rate splitting at the interfering transmitter is derived. Under two partial order broadcast conditions {\em interference-oblivious less noisy} and {\em interference-cognizant less noisy}, a reduced form of $\mathcal{R}$ is shown to be equivalent to the region based on a simpler scheme that uses only superposition coding at the broadcast transmitter. Furthermore, the capacity regions of DM-BIC under the two partial order broadcast conditions are characterized respectively for the strong and very strong interference conditions.
\end{abstract}

\section{Introduction}
Broadcast channel and interference channel are two important classes of multi-user channels that have drawn considerable research attention in the past few decades, mostly due to their simplicity as a fundamental building block and their close relevance to practical communication networks. While complete characterizations are not available, there have been significant advances on these topics in the information theory literature. Notably the best general achievable schemes for the two channels are respectively given by Marton \cite{Marton} and Han-Kobayashi \cite{Han}, which are capacity achieving for some subclass channels or under various conditions, such as the ones in \cite{Cover BC}-\cite{Etkin}.

Motivated by an recent interest in a heterogeneous cellular network design paradigm \cite{Andrews}, we explore a multi-user channel that combines the broadcasting and interference features, i.e. broadcast interference channel (BIC). Specifically we envision a communication scenario where a macro base station (BS) broadcasts to two macro users, one of which is interfered by a point-to-point transmission from a femto BS to a femto user. While the BIC studied presents a simplified version of what might happen in practice, we believe that a fundamental understanding of this simpler channel is crucial for characterizing the trade-offs in heterogeneous networks.

Variations of BIC have been previously studied by Shang and Poor in \cite{Shang I}, for a different interference profile where interference is from the broadcast transmitter to the point-to-point receiver, and in \cite{Shang II}, for the Gaussian BIC where both of the broadcast receivers are subject to interference. Even though the channel studied in this paper has a more restrictive interference profile than that in \cite{Shang II}, we address the more general discrete memoryless channel and provide more general classes of common strategies as well as capacity regions under some conditions. Specifically, we derive an achievable rate region $\mathcal{R}$ based on rate splitting, superposition coding and binning at the broadcast transmitter and rate splitting at the interfering transmitter. This region is a natural generalization of Marton's region \cite{Marton} for a DM-BC. We then define two partial order broadcast conditions, {\em interference-oblivious less noisy} and {\em interference-cognizant less noisy}. Under these conditions, a reduced form of $\mathcal{R}$ is shown to be equivalent to the region based on a simpler scheme that uses only superposition coding at the broadcast transmitter. Furthermore, if interference is strong for the interference-oblivious less noisy DM-BIC, the capacity region is given by the aforementioned two equivalent rate regions. Interestingly, for the interference-cognizant less noisy DM-BIC, we argue that the strong but not very strong interference condition does not exist and in this case, we obtain the capacity region for the very strong interference.

This paper is organized as the follows. The channel model is introduced in Section II, followed by the derivation of $\mathcal{R}$ in Section III. For DM-BIC with two partial order broadcast conditions, the equivalence of rate regions is presented in Section IV and the capacity regions are derived in Section V. This paper is concluded in Section VI.

\emph{Notation}: Let $\phi$ denote a constant. The notation convention follows \cite{Cover}.

\section{Channel Model}
\begin{figure}[htb]
    \centering
    \includegraphics[width=65mm]{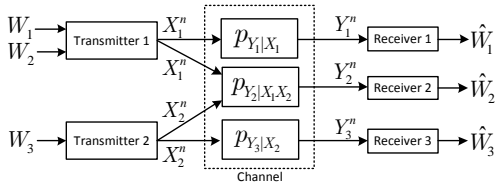}
    \caption{Channel Model}
    \label{ChnlMo}
\end{figure}

A discrete memoryless broadcast interference channel is denoted by $(\mathcal{X}_1\times \mathcal{X}_2 ,p(y_1,y_2,y_3|x_1,x_2), \mathcal{Y}_1\times\mathcal{Y}_2\times\mathcal{Y}_3)$, where $\mathcal{X}_i$, $i=1,2$, are the input alphabets, $\mathcal{Y}_j$, $j=1,2,3$, are the output alphabets and $p(y_1,y_2,y_3|x_1,x_2)$ is the channel transition probability. In this paper, we concentrate on a specific interference profile, where $p(y_1,y_2,y_3|x_1,x_2)= p(y_1|x_1)p(y_2|x_1,x_2)p(y_3|x_2)$. As shown in Fig \ref{ChnlMo}, while transmitter 1 wishes to broadcast to receivers 1, 2 , the second receiver is interfered by transmitter 2 who wishes to communicate with receiver 3.

\begin{definition}
A $(M_1,M_2,M_3,n)$ code consists of message sets $\mathcal{W}_j=\{1,...,M_j\}$; two encoding functions $X_1: (\mathcal{W}_1\times\mathcal{W}_2) \rightarrow \mathcal{X}_1^n$, $X_2:\mathcal{W}_3\rightarrow \mathcal{X}_2^n$ and three decoding functions $g_j:\mathcal{Y}_j^n\rightarrow \mathcal{W}_j$, $j=1,2.3$.
\end{definition}

The messages $W_j$ are uniformly from $\mathcal{W}_j$. The average error probability for the $(M_1,M_2,M_3,n)$ code is
\begin{equation*}
    P_e=\textrm{Pr}(g_1(Y_1^n)\neq W_1 \textrm{ or }g_2(Y_2^n)\neq W_2 \textrm{ or }g_3(Y_3^n)\neq W_3).
\end{equation*}

\begin{definition}
Rates of a $(M_1,M_2,M_3,n)$ code are defined as $R_j=\frac{\log_2(M_j)}{n}$ for $j=1,2,3$.
\end{definition}

Rates $(R_1,R_2,R_3)$ are said to be \textit{achievable} if there exists a sequence of $(M_1,M_2,M_3,n)$ codes with $P_e\rightarrow 0$ as $n\rightarrow \infty$. An achievable rate region is the set of all achievable rates for a given coding scheme. The capacity region is the closure of the union of all achievable rate regions.

\section{Achievable Rate Region for a General DM-BIC}
In this section, we derive an achievable rate region for a general DM-BIC, where the broadcast transmitter employs rate splitting, superposition coding and binning and the interfering transmitter employs rate splitting.

\begin{theorem}
\label{BinningRegion}
$\mathcal{R}$ is an achievable rate region for DM-BIC, where $\mathcal{R}$ is the closure of all $(R_1,R_2,R_3)$ satisfying
{\small
\begin{align}
    R_1&\leq I(V_1;Y_1|Q)\notag\\
    R_2&\leq I(V_2;Y_2|U_2,Q)\notag\\
    R_3&\leq I(X_2;Y_3|Q)\notag\\
    R_1+R_2&\leq I(V_1;Y_1|U_1,Q) + I(V_2;Y_2|U_2,Q)-\notag\\
     &\quad\  I(V_1;V_2|U_1,Q)\notag\\
    R_1+R_2&\leq I(V_1;Y_1|Q) + I(V_2;Y_2|U_1,U_2,Q)-\notag\\
    &\quad\  I(V_1;V_2|U_1,Q)\notag\\
    R_2+R_3&\leq I(V_2,U_2;Y_2|Q)+I(X_2;Y_3|U_2,Q)\label{R2plusR3}\\
    R_1+R_2+R_3&\leq I(V_1;Y_1|U_1,Q)+I(V_2,U_2;Y_2|Q)+\notag\\
    &\quad\ I(X_2;Y_3|U_2,Q)-I(V_1;V_2|U_1,Q)\label{sum1}\\
    R_1+R_2+R_3&\leq I(V_1;Y_1|Q)+I(V_2,U_2;Y_2|U_1,Q)+\notag\\
    &\quad\ I(X_2;Y_3|U_2,Q)-I(V_1;V_2|U_1,Q)\label{sum2}\\
    R_1,R_2,R_3&\geq 0\notag
\end{align}
}for some function $X_1=f(U_1,V_1,V_2)$ and joint distribution
{\small\begin{align*}
    \mathcal{P}&_{Q,U_1,V_1,V_2,U_2,X_2}\\
    &=p(q)p(u_1|q)p(v_1|q,u_1)p(v_2|q,u_1)p(u_2|q)p(x_2|u_2,q),
\end{align*}
}such that
{\small
\begin{align}
    I(V_1;Y_1|U_1,Q)+I(V_2;Y_2|U_1,U_2,Q)-I(V_1;V_2|U_1,Q)\geq 0. \label{cnst1}
\end{align}}
\end{theorem}

\begin{IEEEproof}
The proof is relegated to App. \ref{ProofTheoremBinningRegion}. Here we provide a sketch. The messages for receivers 1, 2 are split into common and private parts respectively. Common messages are carried by the cloud signal $U_1$, which is decoded at both $Y_1$ and $Y_2$. The private message carriers $V_1,V_2$, which are only decoded at their respective intended receivers and treated as noise elsewhere, are superimposed upon $U_1$, where binning is used to allow arbitrary dependence between $V_1$ and $V_2$. At the interfering transmitter, rate splitting is employed to alleviate interference by receiver 2 decoding the common signal $U_2$ while treating the private as noise.
\end{IEEEproof}

\begin{remark}
Constraint (\ref{cnst1}) on the choice of joint input distributions is a direct consequence of the nonnegativity of some intermediate rates, which are eventually eliminated using Fourier-Motzkin procedure.
\end{remark}

\begin{remark}
With $U_1=V_1=\phi$, $X_1=V_2$ and $R_1=0$, $\mathcal{R}$ reduces to the compact Han-Kobayashi region \cite{CMG} for a one-sided interference channel. With $X_2=U_2=\phi$ and $R_3=0$, $\mathcal{R}$ reduces to the most general form of Marton's region with private message sets for a general DM-BC \cite{El Gamal}. Notice that with $U_2=\phi$, the constraint (\ref{cnst1}) reduces to 
\begin{align}
  I(V_1;Y_1|U_1,Q)+I(V_2;Y_2|U_1,Q)-I(V_1;V_2|U_1,Q)\geq 0, \label{eq:ineq}
\end{align}
which applies to Marton's region. However a closer examination reveals that (\ref{eq:ineq}) is unnecessary. For some joint distribution that violates (\ref{eq:ineq}), we have $R_1,R_2\leq I$, where $I=\min\{I(U_1;Y_1),I(U_1;Y_2)\}$. Clearly $(R_1,R_2)$ is contained in Marton's region for some other joint distribution that satisfies (\ref{eq:ineq}). Therefore removing constraint (\ref{eq:ineq}) does not really enlarge the region.
\end{remark}

\section{Equivalence of Rate Regions for the DM-BIC under Partial Order Broadcast Conditions}
Here we concentrate on DM-BIC under two partial order broadcast conditions: interference-oblivious less noisy and interference-cognizant less noisy, which will be defined next.

\begin{definition}
In a DM-BIC, receiver 2 is said to be {\em interference-oblivious less noisy} than receiver 1, denoted by $Y_1\prec_o Y_2$, if $I(U_1;Y_1)\leq I(U_1;Y_2)$ for all $p(u_1,x_1)p(x_2)$ such that $U_1\rightarrow (X_1,X_2)\rightarrow (Y_1,Y_2)$.
\end{definition}

\begin{definition}
In a DM-BIC, receiver 1 is said to be {\em interference-cognizant less noisy} than receiver 2, denoted by $Y_1\succ_c Y_2$, if $I(U_1;Y_1)\geq I(U_1;Y_2|X_2)$ for all $p(u_1,x_1)p(x_2)$ such that $U_1\rightarrow (X_1,X_2)\rightarrow (Y_1,Y_2)$.
\end{definition}

\begin{remark}
We can interpret $Y_1\prec_o Y_2$ as the follows: receiver 2 is {\em less noisy} than receiver 1 \cite{Korner}, even though no particular action is  taken by receiver 2 to deal with interference. Similarly, $Y_1\succ_c Y_2$ can be interpreted as the follows: even if interference $X_2$ is provided to receiver 2, receiver 1 is still less noisy. Also note that degradedness (physical or stochastic, which are the same in broadcast channel \cite[Theorem 15.6.1]{Cover}) implies the partial order conditions and hence is stricter. For example, $Y_2$ being degraded with respect to $Y_1$, i.e. $X_1\rightarrow Y_1\rightarrow Y_2$ holds for all $p(x_1,x_2)$, implies $Y_1\succ_c Y_2$, but not vice versa.
\end{remark}

The first class of schemes we consider is a specialization of $\mathcal{R}$, given in the following two Corollaries.

\begin{corollary}
\label{CoroR1}
$\mathcal{R}_1$ is an achievable rate region for DM-BIC with $Y_1\prec_o Y_2$, where $\mathcal{R}_1$ is the closure of all $(R_1,R_2,R_3)$ satisfying
{\small\begin{align}
    R_1&\leq I(U_1;Y_1)\notag\\
    R_3&\leq I(X_2;Y_3)\notag\\
    R_1+R_2&\leq I(U_1;Y_1)+I(X_1;Y_2|U_1,U_2) \label{largestR22}\\
    R_1+R_2+R_3&\leq I(U_1;Y_1)+I(X_1,U_2;Y_2|U_1)+\notag \\
    &\quad\ I(X_2;Y_3|U_2) \label{largestR33}\\
    R_1,R_2,R_3&\geq 0\notag
\end{align}
}for some $\mathcal{P}_{U_1,X_1,U_2,X_2}=p(u_1)p(x_1|u_1)p(u_2)p(x_2|u_2)$.
\end{corollary}

\begin{IEEEproof}
Fix $Q=\phi$. Specializing $\mathcal{R}$ with $X_1=V_2$, $V_1=U_1$ and removing redundant inequalities due to $Y_1\prec_o Y_2$, we obtain $\mathcal{R}_1$.
\end{IEEEproof}

\begin{corollary}
\label{CoroR2}
$\mathcal{R}_2$ is an achievable rate region for DM-BIC with $Y_1\succ_c Y_2$, where $\mathcal{R}_2$ is the closure of all $(R_1,R_2,R_3)$ satisfying
{\small\begin{align}
    R_2&\leq I(U_1;Y_2|U_2)\label{largestR2}\\
    R_3&\leq I(X_2;Y_3)\notag\\
    R_1+R_2&\leq I(X_1;Y_1|U_1)+I(U_1;Y_2|U_2)\label{largestR1}\\
    R_2+R_3&\leq I(U_1,U_2;Y_2)+I(X_2;Y_3|U_2)\label{largestR3}\\
    R_1+R_2+R_3&\leq I(X_1;Y_1|U_1)+I(U_1,U_2;Y_2)+I(X_2;Y_3|U_2)\notag\\
    R_1,R_2,R_3&\geq 0\notag
\end{align}
}for some $\mathcal{P}_{U_1,X_1,U_2,X_2}=p(u_1)p(x_1|u_1)p(u_2)p(x_2|u_2)$.
\end{corollary}

\begin{IEEEproof}
A direct specialization of $\mathcal{R}$ will result in some extra inequalities that are harder to remove. Hence we take an indirect approach, where the specialization is done for an equivalent region of $\mathcal{R}$. The details are provided in App. \ref{ProofCoroR2}.
\end{IEEEproof}

Notice that to derive $\mathcal{R}_i$, $i=1,2$, we fix the time-sharing r.v. $Q$. In principle, we could have kept $Q$ intact when specializing $\mathcal{R}$, but the following proposition asserts that there is no benefit doing so. Since time-sharing always results in a region no smaller than convex hull operation, it follows that taking convex hull is also unnecessary.

\begin{proposition}
\label{convexification}
Time-sharing does not enlarge $\mathcal{R}_i$, $i=1,2$.
\end{proposition}

\begin{IEEEproof}
The proof is relegated to App. \ref{ProofPropositionConv}.
\end{IEEEproof}

Next we present two achievable rate regions, $\mathcal{R}_{(i)}$, $i=1,2$, which are solely based on superposition coding (i.e. no rate splitting at the broadcast transmitter), where the cloud center carries only receiver $i$'s message. Since the proofs are standard, they are omitted for conciseness.

\begin{theorem}
\label{Rprime}
$\mathcal{R}_{(1)}$ is an achievable rate region for DM-BIC with $Y_1\prec_o Y_2$, where $\mathcal{R}_{(1)}$ is the closure of all $(R_1,R_2,R_3)$ satisfying
{\small\begin{align*}
    R_1&\leq I(U_1;Y_1)\\ 
    R_2&\leq I(X_1;Y_2|U_1,U_2)\\
    R_3&\leq I(X_2;Y_3)\\
    R_2+R_3&\leq I(X_1,U_2;Y_2|U_1)+I(X_2;Y_3|U_2)\\
    R_1,R_2,R_3&\geq 0 
\end{align*}
}for some $\mathcal{P}_{U_1,X_1,U_2,X_2}=p(u_1)p(x_1|u_1)p(u_2)p(x_2|u_2)$.
\end{theorem}

\begin{theorem}
\label{Rpprime}
$\mathcal{R}_{(2)}$ is an achievable rate region for DM-BIC with $Y_1\succ_c Y_2$, where $\mathcal{R}_{(2)}$ is the closure of all $(R_1,R_2,R_3)$ satisfying
{\small\begin{align*}
    R_1&\leq I(X_1;Y_1|U_1)\\
    R_2&\leq I(U_1;Y_2|U_2)\\
    R_3&\leq I(X_2;Y_3)\\
    R_2+R_3&\leq I(U_1,U_2;Y_2)+I(X_2;Y_3|U_2)\\
    R_1,R_2,R_3&\geq 0
\end{align*}
}for some $\mathcal{P}_{U_1,X_1,U_2,X_2}=p(u_1)p(x_1|u_1)p(u_2)p(x_2|u_2)$.
\end{theorem}

In deriving the most general region $\mathcal{R}$, we used rate splitting, superposition coding and binning at the broadcast transmitter. Regions $\mathcal{R}_i$, $i=1,2$, are derived from $\mathcal{R}$ when binning is stripped off but rate splitting and superposition kept intact. While both $\mathcal{R}_i$ and $\mathcal{R}_{(i)}$ rely on superposition coding, there is a subtle difference. Despite the fact that both schemes' cloud centers carry receiver $i$'s message, the one for $\mathcal{R}_i$ could also carry receiver $j$'s ($j\neq i$, $j=1,2$) common message, which could be potentially helpful to reduce the self-interference due to the fact that part of the broadcast signal intended for receiver $j$ is essentially interference from receiver $i$'s perspective. It is apparent that the superposition-only-based rate regions are not larger than the ones based on superposition and rate splitting, since the latter includes the former as a special case. This can be also verified by explicitly checking that the inequalities defining $\mathcal{R}_{(i)}$ induce those in $\mathcal{R}_i$, but not vice versa. Hence at first sight it seems that $\mathcal{R}_i$ is strictly larger than $\mathcal{R}_{(i)}$. However, if we consider the no interference case, i.e. $U_2=X_2=\phi$, $R_3=0$, $\mathcal{R}_i$ cannot be strictly larger than $\mathcal{R}_{(i)}$ since the latter is the capacity region of a less noisy (or degraded) DM-BC \cite{El Gamal}. The pitfall of the previous argument is that it only considers a specific input distribution. It is true that for some given $\mathcal{P}_{U_1,X_1,U_2,X_2}$, $\mathcal{R}_i$ is strictly larger, however once we consider all $\mathcal{P}_{U_1,X_1,U_2,X_2}$, we will show that they are indeed equivalent.

\begin{theorem}
\label{theoremequi}
$\mathcal{R}_i=\mathcal{R}_{(i)}$, $i=1,2$.
\end{theorem}

\subsection{Proof of Theorem \ref{theoremequi}}
Before proving Theorem \ref{theoremequi}, we need the following definitions and lemmas.

\begin{definition}
Let $\mathbb{R}_c^n$ be a convex subset of $\mathbb{R}^n$, a $n$-dimensional Euclidean space. A point $X\in\mathbb{R}_c^n$ is an \textit{extreme point (ExP)} iff whenever $X=tY+(1-t)Z$, $t\in(0,1)$ and $Y\neq Z$, this implies either $Y\not \in \mathbb{R}_c^n$ or $Z\not \in \mathbb{R}_c^n$.
\end{definition}

\begin{definition}
An ExP $X\in\mathbb{R}_c^n$ is said to be \textit{dominant (DExP)} iff there does not exist another ExP $Y\in\mathbb{R}_c^n$, $Y\neq X$, such that $X\leq Y$ element-wise.
\end{definition}

\begin{remark}
In the literature, the term ``dominant extreme points'' are sometimes referred as corner points. The intention of choosing the former terminology is to emphasize the connection to convex set.
\end{remark}

Let $\mathcal{R}^n$ be a $n$-dimensional convex rate region, of which the set of all DExPs is denoted by $\Omega$. Further let $co(\Omega)$ denote the convex hull of $\Omega$:
{\small\begin{align*}
  co(\Omega)=\left\{\sum_{i=1}^{m}\alpha_i\mathbf{R}_i \left|\mathbf{R}_i \in\Omega, \alpha_i\in[0,1], \sum_{i=1}^m\alpha_i=1, m=1,2,... \right.\right\}.
\end{align*}

}
\begin{lemma}
\label{DominateEx}
$\mathbf{R}\in\mathcal{R}^n$ iff there exists some $\mathbf{R}'\in co(\Omega)$ such that $\mathbf{R}\leq \mathbf{R}'$ element-wise.
\end{lemma}

\begin{IEEEproof}
For the ``if'' part, since DExPs are achievable, so is their convex combination, specifically $\mathbf{R}'$ is achievable. If $\mathbf{R}\leq \mathbf{R}'$, $\mathbf{R}$ is also achievable. For the ``only if'' part, for a convex region $\mathcal{R}^n$, any achievable rate can be expressed as a convex combination of some ExPs, i.e. there exists some $\mathbf{R}_i \in\Psi$, $\alpha_i\in[0,1]$ and an integer $m$ such that $\mathbf{R}=\sum_{i=1}^m \alpha_i\mathbf{R}_i$, where $\Psi$ denotes the set of all ExPs for $\mathcal{R}^n$. Now replacing any non-dominant $\mathbf{R}_i$ that constitutes $\mathbf{R}$ by its corresponding DExP and keeping convex coefficients $\alpha_i$ intact, we obtain $\mathbf{R}'\in co(\Omega)$, where $\mathbf{R}\leq \mathbf{R}'$.
\end{IEEEproof}

Lemma \ref{DominateEx} suggests that a rate region is completely described by its DExPs. When comparing different rate regions, it suffices to consider their sets of DExPs, which will be given in the follows for $\mathcal{R}_i$ and $\mathcal{R}_{(i)}$, $i=1,2$, respectively.

\begin{lemma}
\label{DEXPR2minus}
For a $\mathcal{P}_{U_1,X_1,U_2,X_2}$, DExPs of $\mathcal{R}_{2}$ are given by:
{\footnotesize
\begin{align*}
    A &= \Big( I(X_1;Y_1|U_1),\ I(U_1;Y_2|U_2),\ \min\{ I(X_2;Y_3),\ I(U_2;Y_2)+\\
    &\qquad I(X_2;Y_3|U_2) \}  \Big)\\
    B &= \Big( I(X_1;Y_1|U_1),\ \min\{I(U_1;Y_2|U_2),\ [I(U_1,U_2;Y_2)-I(U_2;Y_3)]^+ \} ,\\
    &\quad\quad  I(X_2;Y_3|U_2)+\min\{ I(U_2;Y_3) ,\ I(U_1,U_2;Y_2) \}  \Big)\\
    C &= \Big( I(X_1;Y_1|U_1)+ \min\{I(U_1;Y_2|U_2),\ [I(U_1,U_2;Y_2)-I(U_2;Y_3)]^+ \},\\
     &\qquad 0 ,\ I(X_2;Y_3|U_2)+\min\{ I(U_2;Y_3) ,\ I(U_1,U_2;Y_2) \}  \Big)\\
    D &= \Big( I(X_1;Y_1|U_1)+I(U_1;Y_2|U_2),\ 0 ,\ \min\{ I(X_2;Y_3) ,\  I(U_2;Y_2)+\\
    &\qquad I(X_2;Y_3|U_2) \}  \Big)
\end{align*}}
\end{lemma}

\begin{IEEEproof}
The proof is relegated to App. \ref{proofDEXPR2minus}.
\end{IEEEproof}

\begin{lemma}
\label{DEXPRpprimeminus}
For a $\mathcal{P}_{U_1,X_1,U_2,X_2}$, $\mathcal{R}_{(2)}$ has two DExPs $A$, $B$ as in Lemma \ref{DEXPR2minus}.
\end{lemma}

\begin{IEEEproof}
The proof is relegated to App. \ref{proofDEXPRpprimeminus}.
\end{IEEEproof}

\begin{lemma}
\label{DEXPR1minus}
For a $\mathcal{P}_{U_1,X_1,U_2,X_2}$, DExPs of $\mathcal{R}_{1}$ include
{\footnotesize
\begin{align*}
    E&= \Big( I(U_1;Y_1),\ I(X_1;Y_2|U_1,U_2),\ \min\{ I(X_2;Y_3),\ I(U_2;Y_2|U_1)+\\
    &\quad\quad I(X_2;Y_3|U_2) \}  \Big)\\
    F&=\Big( I(U_1;Y_1),\ \min\{ I(X_1;Y_2|U_1,U_2),\ [I(X_1,U_2;Y_2|U_1)-\\
    &\qquad I(U_2;Y_3)]^+\} ,\ I(X_2;Y_3|U_2) + \min\{ I(U_2;Y_3) ,\\
     &\qquad  I(X_1,U_2;Y_2|U_1) \}\Big)\\
    G&=\Big( 0,\ I(U_1;Y_1)+I(X_1;Y_2|U_1,U_2),\ \min\{ I(X_2;Y_3) ,\ I(U_2;Y_2|U_1)\\
    &\quad\ \ +I(X_2;Y_3|U_2)\} \Big).
\end{align*}
}Furthermore, if $\mathcal{P}_{U_1,X_1,U_2,X_2}$ satisfies $I(U_2;Y_3)\leq I(U_1;Y_1)+I(X_1,U_2;Y_2|U_1)$, there are two more DExPs
{\footnotesize
\begin{align*}
    H &=\Big( 0,\ I(U_1;Y_1)+ I(X_1;Y_2|U_1,U_2)+\min\{0,\ I(U_2;Y_2|U_1)\\
    &\qquad -I(U_2;Y_3)\},\  I(X_2;Y_3)  \Big)\\
    I &= \Big( I(U_1;Y_1) + \min\{ 0 ,\ I(X_1,U_2;Y_2|U_1) - I(U_2;Y_3)\} \},\\
    &\qquad [I(X_1;Y_2|U_1,U_2)+\min\{ 0 ,\ I(U_2;Y_2|U_1)-I(U_2;Y_3)\}]^+,\\
    &\qquad I(X_2;Y_3) \}   \Big).
\end{align*}
}Otherwise if $I(U_2;Y_3)> I(U_1;Y_1)+I(X_1,U_2;Y_2|U_1)$, there is one more DExP
{\footnotesize
\begin{align*}
    J=\Big(0,\ 0,\ I(U_1;Y_1)+I(X_1,U_2;Y_2|U_1)+I(X_2;Y_3|U_2) \Big).
\end{align*}
}
\end{lemma}

\begin{IEEEproof}
The proof is relegated to App. \ref{proofDEXPR1minus}.
\end{IEEEproof}

\begin{lemma}
For a $\mathcal{P}_{U_1,X_1,U_2,X_2}$, $\mathcal{R}_{(1)}$ has two DExPs $E$, $F$ as in Lemma \ref{DEXPR1minus}.
\end{lemma}

\begin{IEEEproof}
The proof is exactly the same as Case 1 for $\mathcal{R}_1$.
\end{IEEEproof}

\begin{IEEEproof}[Proof of Theorem \ref{theoremequi}]
We will first prove $\mathcal{R}_{2}=\mathcal{R}_{(2)}$ and then $\mathcal{R}_{1}=\mathcal{R}_{(1)}$. Let $\mathcal{P}=\mathcal{P}_{U_1,X_1,U_2,X_2}$. We use $\mathcal{P}_{U_i=\phi}$ to denote the same distribution except that $U_i=\phi$.

\noindent \textit{1. Proof of $\mathcal{R}_{2}=\mathcal{R}_{(2)}$}

From Lemma \ref{DEXPR2minus} and \ref{DEXPRpprimeminus}, for a given $\mathcal{P}$, $\mathcal{R}_{2}$ has two more DExPs than $\mathcal{R}_{(2)}$. However, for $\mathcal{P}_{U_1=U_2=\phi}$, $A$ becomes $A'=\left(\ I(X_1;Y_1),\ 0,\ I(X_2;Y_3)\ \right)$ and it can be shown $C,D\leq A'$ due to $Y_1\succ_c Y_2$. Therefore if we take the union of regions for the two distributions $\mathcal{P}$ and $\mathcal{P}_{U_1=U_2=\phi}$, both $\mathcal{R}_{2}$ and $\mathcal{R}_{(2)}$ will have identical DExPs. By Lemma \ref{DominateEx}, $\mathcal{R}_{2}=\mathcal{R}_{(2)}$.

\noindent \textit{2. Proof of $\mathcal{R}_{1}=\mathcal{R}_{(1)}$}

This part is more involved, but the idea is essentially the same. We first show $G$ is redundant. Given any $\mathcal{P}$, consider another joint distribution $\mathcal{P}_{U_1=\phi}$. Then $E$ and $F$ become
{\footnotesize\begin{align*}
    E' &= \Big(\ 0,\ I(X_1;Y_2|U_2),\ \min\{ I(X_2;Y_3),\ I(U_2;Y_2)+I(X_2;Y_3|U_2) \}  \Big)\\
    F' &= \Big(\ 0,\ \min\{I(X_1;Y_2|U_2),\ [I(X_1,U_2;Y_2)-I(U_2;Y_3)]^+ \} ,\\
     &\qquad\ I(X_2;Y_3|U_2)+\min\{ I(U_2;Y_3) ,\ I(X_1,U_2;Y_2) \} \Big).
\end{align*}
}The region, specified by DExPs $E'$ and $F'$, can be alternatively described by the following inequalities (this can be verified by setting $U_1=\phi$ in $\mathcal{R}_{1}$)
{\small\begin{align*}
    R_3&\leq I(X_2;Y_3)\\
    R_2&\leq I(X_1;Y_2|U_2)\\
    R_2+R_3&\leq I(X_1,U_2;Y_2)+I(X_2;Y_3|U_2)\\
    R_2,R_3&\geq 0, \ R_1=0
\end{align*}
}Using the fact $I(U_1;Y_1)\leq I(U_1;Y_2)$ due to $Y_1\prec_o Y_2$, it can be checked that $G$ is contained in the above region. Hence $G$ is redundant.

Next we will show that $H$, $I$ either reduce to other DExPs or are redundant if we consider all input distributions. Let us first focus on $H$. If $I(U_2;Y_3)< I(U_2;Y_2|U_1)$, $G=\big(0,\ I(U_1;Y_1)+I(X_1;Y_2|U_1,U_2),\ I(X_2;Y_3)\big)$ and $H= G$. If $\delta: \ I(U_2;Y_2|U_1) \leq I(U_2;Y_3) < I(X_1,U_2;Y_2|U_1)$, we have
{\small\begin{align*}
    F&=\Big( I(U_1;Y_1),\ I(X_1,U_2;Y_2|U_1)-I(U_2;Y_3),\ I(X_2;Y_3) \Big)\\
    H&=\Big( 0,\ I(U_1;Y_1)+ I(X_1,U_2;Y_2|U_1)-I(U_2;Y_3),\ I(X_2;Y_3)  \Big)
\end{align*}
}Notice that for any $\mathcal{P}$, if condition $\delta$ holds, then for $\mathcal{P}_{U_1=\phi}$, $\delta$ still holds. Hence if we let $U_1=\phi$, $F$ becomes $F'=\big( 0,\ I(X_1,U_2;Y_2)-I(U_2;Y_3),\ I(X_2;Y_3) \big)$ and $H\leq F'$ due to $Y_1\prec_o Y_2$. If $I(X_1,U_2;Y_2|U_1) \leq I(U_2;Y_3) \leq I(U_1;Y_1)+I(X_1,U_2;Y_2|U_1)$, by setting $U_2=\phi$ we have
{\small\begin{align*}
    G'=\Big( 0,\ I(U_1;Y_1)+I(X_1;Y_2|U_1),\ I(X_2;Y_3) \Big)\geq H.
\end{align*}

}
We now consider $I$. If $I(U_2;Y_3)\leq I(U_2;Y_2|U_1)$, $I=E$. If $I(U_2;Y_2|U_1)<I(U_2;Y_3)< I(X_1,U_2;Y_2|U_1)$, $I=F$. If $I(X_1,U_2;Y_2|U_1)\leq I(U_2;Y_3)\leq I(U_1;Y_1)+I(X_1,U_2;Y_2|U_1)$, $I$ reduces to
{\small\begin{align*}
    I=\Big( I(U_1;Y_1)+ I(X_1,U_2;Y_2|U_1)-I(U_2;Y_3),\ 0,\ I(X_2;Y_3) \Big)
\end{align*}
}If we further let $U_1=X_1$ and $U_2=\phi$, $E$ becomes $E'=\big( I(X_1;Y_1),\ 0,\ I(X_2;Y_3)\big)$. Clearly, $I\leq E'$.

At last, we consider $J$. For any $\mathcal{P}$, setting $U_1=U_2=\phi$, $E$ becomes
{\small\begin{align*}
    E'=\Big( 0,\ I(X_1;Y_2),\ I(X_2;Y_3) \Big).
\end{align*}
}Clearly, $J<E'$ due to the condition $I(U_1;Y_1)+I(X_1,U_2;Y_2|U_1)< I(U_2;Y_3)$ and hence $J$ is redundant.

To summarize, even though for a specific $\mathcal{P}$, $\mathcal{R}_{1}$ could have more DExPs than $\mathcal{R}_{(1)}$, if we consider all possible $\mathcal{P}$, they will have exactly the same set of DExPs given by $E$ and $F$. By Lemma \ref{DominateEx}, $\mathcal{R}_{1}=\mathcal{R}_{(1)}$.
\end{IEEEproof}

\section{Capacity Regions under the Strong/Very Strong Interference Condition}
In this section, capacity regions of DM-BIC with $Y_1\prec_o Y_2$ and $Y_1\succ_c Y_2$ are established respectively for the strong and very strong interference conditions defined in the following.

\begin{definition}
\label{strongcondition}
Interference is said to be {\em strong} if for all $p(x_1)p(x_2)$, $I(X_2;Y_2|X_1)\geq I(X_2;Y_3)$.
\end{definition}
\begin{definition}
\label{verystrongcondition}
Interference is said to be {\em very strong} if for all $p(x_1)p(x_2)$, $I(X_2;Y_2)\geq I(X_2;Y_3)$.
\end{definition}

\begin{remark}
The intuition behind these definitions, which are the same as the regular interference channel \cite{Costa}, is that by conditioning on the intended signal, whose decoding is assured to be successful by design, the interfered receiver sees a better channel than interference's own receiver. This suggests that the interfered receiver should be able to decode the interference along with its intended signal, by performing a joint decoding if interference is strong. If further interference is very strong, successive interference cancellation decoding suffices, where interference is decoded first. Evidently very strong condition is stricter than the strong condition.
\end{remark}

\begin{theorem}
\label{capastrong}
The capacity region of DM-BIC with $Y_1\prec_o Y_2$ and the strong interference condition is the closure of all $(R_1,R_2,R_3)$ satisfying
{\small\begin{align*}
    R_1&\leq I(U_1;Y_1)\\
    R_2&\leq I(X_1;Y_2|U_1,X_2)\\
    R_3&\leq I(X_2;Y_3)\\
    R_2+R_3&\leq I(X_1,X_2;Y_2|U_1)\\
    R_1,R_2,R_3&\geq 0
\end{align*}
}for some $\mathcal{P}_{U_1,X_1,X_2}=p(u_1)p(x_1|u_1)p(x_2)$.
\end{theorem}

\begin{IEEEproof}
The proof is relegated to App. \ref{proofcapastrong}.
\end{IEEEproof}

\begin{remark}
The capacity region takes two different forms. The one given in Theorem \ref{capastrong} is identical to $\mathcal{R}_{(1)}$ with $U_2=X_2$. An alternative form is given by $\mathcal{R}_{1}$ with $U_2=X_2$.
\end{remark}

When receiver 2 is interference-oblivious less noisy than receiver 1, for any sensible coding scheme $X_1$ should always be decodable at receiver 2 (otherwise, none of the broadcast receivers can do so). Hence the strong condition, originated from interference channel, naturally carries over to DM-BIC with $Y_1\prec_o Y_2$. However, this is not the case for DM-BIC with $Y_1\succ_c Y_2$, which will be discussed next.

\begin{theorem}
\label{capaverystrong}
The capacity region of DM-BIC with $Y_1\succ_c Y_2$ and the very strong interference condition is the closure of all $(R_1,R_2,R_3)$ satisfying
{\small\begin{align*}
    R_1&\leq I(X_1;Y_1|U_1)\\
    R_2&\leq I(U_1;Y_2|X_2)\\
    R_3&\leq I(X_2;Y_3)\\
    R_1,R_2,R_3&\geq 0
\end{align*}
}for some $\mathcal{P}_{U_1,X_1,X_2}=p(u_1)p(x_1|u_1)p(x_2)$.
\end{theorem}

\begin{IEEEproof}
The achievability follows those for $\mathcal{R}_{(2)}$ and $\mathcal{R}_{2}$, all with with $U_2=X_2$. The converse proof is standard.
\end{IEEEproof}

\begin{remark}
Similarly to Theorem \ref{capastrong}, the capacity region takes two forms, one given in Theorem \ref{capaverystrong}, which is essentially $\mathcal{R}_{(2)}$ with $U_2=X_2$, and another given by $\mathcal{R}_{2}$ with $U_2=X_2$.
\end{remark}

It is not difficult to see that strong condition in Definition \ref{strongcondition} does not fit well for DM-BIC with $Y_1\succ_c Y_2$. The reason is that if $X_1$ is the intended signal for receiver 2, i.e. $X_1$ always decodable at receiver 2, then by $Y_1\succ_c Y_2$, receiver 1 can decode it as well. Hence the two receivers will always decode the same set of messages, which clearly does not represent the most general case. In fact, we claim that the strong but not very strong interference condition does not exist for DM-BIC with $Y_1\succ_c Y_2$. The argument is as the follows.

The problem is to figure out what is the intended signal for receiver 2. Once we find out such a signal, we can mimic the strong condition in Definition \ref{strongcondition}, with modification of conditioning on that signal instead of $X_1$. Suppose there exists some strong condition, then interference $X_2$ is required to be decoded at receiver 2. Under this restriction, we have an upper bound $n(R_2+R_3-\epsilon_n)\leq I(W_2,W_3;Y_2^n)$. Along with other straightforward upper bounds, by the same technique that we used above to prove Theorem \ref{capastrong}, we can show that $\mathcal{R}_{(2)}$ with $U_2=X_2$ is the capacity region. This implies that if there exists some strong condition, then superposition coding with cloud center $U_1$ carrying receiver 2's message is capacity achieving. Hence without loss of generality, we can view the cloud center $U_1$ as the intended signal for receiver 2, which in return gives us the strong condition $I(X_2;Y_2|U_1)\geq I(X_2;Y_3)$, for all $p(u_1)p(x_1|u_1)p(x_2)$ such that $U_1\rightarrow (X_1,X_2)\rightarrow (Y_2,Y_3)$ form a Markov chain. However, this condition always implies the very strong condition (consider $U_1=\phi$) and furthermore the strong interference capacity region, $\mathcal{R}_{(2)}$ with $U_2=X_2$, always reduces to the very strong capacity region given in Theorem \ref{capaverystrong}. In other words for $Y_1\succ_c Y_2$, if interference is strong, then it has to be very strong.

\section{Conclusion}
In this paper, we devise a coding scheme combining rate splitting, superposition coding and binning for a general DM-BIC. The obtained achievable rate region is then specialized to DM-BIC under two partial order broadcast conditions: interference-oblivious less noisy and interference-cognizant less noisy. By carefully inspecting the dominant extreme points, the specialized rate region is shown to be equivalent to that based on a simpler scheme that uses only superposition coding at the broadcast transmitter. For the interference-oblivious less noisy DM-BIC, if interference is strong, the capacity region is given by the aforementioned two equivalent rate regions. For the interference-cognizant less noisy DM-BIC, we argue that the strong but not very strong interference condition does not exist and in this case, we obtain the capacity region for very strong interference.

\appendices
\section{Proof of Theorem \ref{BinningRegion}}
\label{ProofTheoremBinningRegion}
We will first obtain an achievable rate region $\hat{\mathcal{R}}$ in Lemma \ref{R3hat}. Then we prove $\mathcal{R}=\hat{\mathcal{R}}$.
\begin{lemma}
\label{R3hat}
$\hat{\mathcal{R}}$ is an achievable rate region for DM-BIC, where $\hat{\mathcal{R}}$ is the closure of all $(R_1,R_2,R_3)$ satisfying all inequalities defining $\mathcal{R}$ plus two more constraints
{\small
\begin{align}
    R_3&\leq I(V_2,U_2;Y_2|U_1,Q) + I(X_2;Y_3|U_2,Q)\label{rdt1}\\
    R_3&\leq I(V_1;Y_1|U_1,Q) + I(V_2,U_2;Y_2|U_1,Q) + I(X_2;Y_3|U_2,Q)-\notag\\
     &\quad\ I(V_1;V_2|U_1,Q)\label{rdt2}.
\end{align}
}
\end{lemma}

\begin{IEEEproof}

\textit{Codebook generation}:

Split $Y_1$'s message into two parts: $m_1$ and $i$. Similarly for $Y_2$, $m_2$ and $j$. Generate $2^{n(R_{1c}+R_{2c})}$ independent codewords $u_1^n(m_1,m_2)$ with each symbol i.i.d according to $p_{U_1}(\cdot)$, $m_1\in\{1,2,...,2^{nR_{1c}}\}$, $ m_2\in\{1,2,...,2^{nR_{2c}}\}$. For each $u_1^n(m_1,m_2)$ generate $2^{n(R_{1p}+R_1')}$ conditionally independent codewords $v_1^n(m_1,m_2,i,i')$ with each symbol i.i.d according to $p_{V_1|U_1}(\cdot|u_1(m_1,m_2))$, $i\in\{1,2,...,2^{nR_{1p}}\}$, $ i'\in\{1,2,...,2^{nR_1'}\}$. Similarly for each $u_1^n(m_1,m_2)$, generate $2^{n(R_{2p}+R_2')}$ conditionally independent codewords $v_2^n(m_1,m_2,j,j')$ with each symbol i.i.d according to $p_{V_2|U_1}(\cdot|u_1(m_1,m_2))$, $j\in\{1,2,...,2^{nR_{2p}}\}$, $ j'\in\{1,2,...,2^{nR_2'}\}$.

Split $Y_3$'s message into two parts: $k$ and $l$. Generate $2^{nT_3}$ independent codewords $u_2^n(k)$ with each symbol i.i.d according to $p_{U_2}(\cdot)$, $k\in\{1,2,...,2^{nT_3}\}$. For each $u_2^n(k)$ generate $2^{nS_3}$ conditionally independent codewords $x_2^n(k,l)$ with each symbol i.i.d according to $p_{X_2|U_2}(\cdot|u_2(k))$, $l\in\{1,2,...,2^{nS_3}\}$.

\textit{Encoding}:

Given message quadruple  $(m_1,i,m_2,j)$, broadcast transmitter tries to find a pair $(i',j')$ such that
{\small\begin{align*}
    (v_1^n(m_1,m_2,i,i'),v_2^n(m_1,m_2,j,j'))\in A_{\epsilon}^{(n)}(V_1,V_2).
\end{align*}
}If there is one or more such pairs, choose one and send $x_1^n=f^n(u_1^n(m_1,m_2),v_1^n(m_1,m_2,i,i'),v_2^n(m_1,m_2,j,j'))$, where $f(\cdot)$ is a deterministic function. If there is no such pair, an error is declared and a predefined codeword is sent. Interference transmitter sends codeword $x_2^n(k,l)$ for message pair $(k,l)$.

Without loss of generality, in the following we assume $(m_1,i,m_2,j,k,l)=(1,1,1,1,1,1)$ is sent.

\textit{Decoding}:

Receiver $Y_1$ looks for $(\hat{m}_1,\hat{m}_2,\hat{i},\hat{i}')$ such that
{\small\begin{align*}
(u_1^n(\hat{m}_1,\hat{m}_2) ,v_1^n(\hat{m}_1,\hat{m}_2,\hat{i},\hat{i}'), y_1^n)\in A_{\epsilon}^{(n)}(U_1,V_1,Y_1).
\end{align*}
}If there is no such quadruple or some such quadruple with either $\hat{m}_1\neq 1$ or $\hat{i}\neq 1$ or both, an error is declared.

Receiver $Y_2$ looks for $(\hat{m}_1,\hat{m}_2,\hat{j},\hat{j}',\hat{k})$ such that
{\small
\begin{align*}
    (u_1^n(\hat{m}_1,\hat{m}_2), v_2^n(\hat{m}_1,\hat{m}_2,\hat{j},\hat{j}'), u_2^n(\hat{k}),y_2^n)\in A_{\epsilon}^{(n)}(U_1,V_2,U_2,Y_2).
\end{align*}
}If there is no such tuple or some such tuple with either $\hat{m}_2\neq 1$ or $\hat{j}\neq 1$ or both, an error is declared.

Receiver $Y_3$ looks for unique $(\hat{k},\hat{l})$ such that
{\small\begin{align*}
    (u_2^n(\hat{k}),x_2^n(\hat{k},\hat{l}),y_3^n)\in A_{\epsilon}^{(n)}(U_2,X_2,Y_2).
\end{align*}
}If there is none or more than one such pair, an error is declared.

\textit{Analysis of error probability}:

\textit{At broadcast encoder}:
Given $(m_1,m_2,i,j)$, with high probability there is at least one $(i',j')$ pair such that $(v_1^n(m_1,m_2,i,i'),v_2^n(m_1,m_2,j,j'))$ is jointly typical if $R_1'+R_2'>I(V_1;V_2|U_1)$ due to mutual covering lemma \cite{El Gamal}.

\textit{At receiver $Y_1$}:
Using standard techniques from \cite{El Gamal}, where all error events are first determined using a joint pmf factorization table and then analyzed individually using packing lemma, it can be shown that the error probability at receiver $Y_1$ can be made arbitrarily small if
{\small\begin{align*}
    R_{1p}+R_1'&\leq I(V_1;Y_1|U_1)\\
    R_{1c}+R_{2c}+R_{1p}+R_1'&\leq I(V_1;Y_1).
\end{align*}

}
\textit{At receiver $Y_2$}:
Similarly it can be shown that the error probability at receiver 2 can be made arbitrarily small if
{\small\begin{align*}
    R_{2p}+R_2'&\leq I(V_2;Y_2|U_1,U_2) \\
    R_{2p}+R_2'+T_3&\leq I(V_2,U_2;Y_2|U_1)\\
    R_{1c}+R_{2c}+R_{2p}+R_2' &\leq I(V_2;Y_2|U_2)\\
    R_{1c}+R_{2c}+R_{2p}+R_2'+T_3 &\leq I(V_2,U_2;Y_2).
\end{align*}

}
\textit{At receiver $Y_3$}:
Similarly it can be shown that the error probability at receiver 3 can be made arbitrarily small if
{\small\begin{align*}
    S_3&\leq I(X_2;Y_3|U_2) \\
    T_3+S_3&\leq I(X_2;Y_3).
\end{align*}

}
Collecting all inequalities, applying Fourier-Motzkin elimination with $R_1=R_{1c}+R_{1p}$, $R_2=R_{2c}+R_{2p}$ and $R_3=T_3+S_3$, and finally including a time sharing variable, we obtain an achievable rate region $\hat{\mathcal{R}}$.
\end{IEEEproof}

\begin{proposition}
\label{equivalence}
Inequalities (\ref{rdt1}) and (\ref{rdt2}) are redundant. Therefore $\mathcal{R}=\hat{\mathcal{R}}$.
\end{proposition}

\begin{IEEEproof}
To prove that (\ref{rdt1}) and (\ref{rdt2}) are redundant, we follow the argument used in \cite{CMG} to simply the Han-Kobayashi region. Fix time-sharing r.v. $Q$. Denote $\mathcal{P}=\mathcal{P}_{U_1,V_1,V_2,U_2,X_2}$.

We first prove that (\ref{rdt1}) is redundant. For a given $\mathcal{P}$, we show that if rate triple $(R_1,R_2,R_3)$ satisfies all inequalities in $\hat{\mathcal{R}}$ except (\ref{rdt1}), then $(R_1,R_2,R_3)\in\hat{\mathcal{R}}_{\mathcal{P}_{U_2=\phi}}$. Hence by time-sharing, (\ref{rdt1}) is redundant.

If (\ref{rdt1}) is violated, we have
{\small\begin{align}
    R_3> I(V_2,U_2;Y_2|U_1) + I(X_2;Y_3|U_2) \label{violated1}
\end{align}
}If $(R_1,R_2,R_3)$ satisfies all inequalities in $\hat{\mathcal{R}}$ except (\ref{rdt1}), then it can be shown that $(R_1,R_2,R_3)$ satisfies the following
{\small\begin{align}
    R_1&\leq I(V_1;Y_1)\notag\\
    R_2&\leq I(V_2;Y_2)\label{equi1}\\
    R_3&\leq I(X_2;Y_3)\notag\\
    R_1+R_2&\leq I(V_1;Y_1|U_1) + I(V_2;Y_2) - I(V_1;V_2|U_1)\label{equi2}\\
    R_1+R_2&\leq I(V_1;Y_1) + I(V_2;Y_2|U_1) - I(V_1;V_2|U_1)\label{equi3},
\end{align}
}where (\ref{equi1}) is obtained from (\ref{R2plusR3}) and (\ref{violated1}), (\ref{equi2}) from (\ref{sum1}) and (\ref{violated1}), (\ref{equi3}) from (\ref{sum2}) and (\ref{violated1}). Notice that the above region is exactly $\hat{\mathcal{R}}_{\mathcal{P}_{U_2=\phi}}$. Hence (\ref{rdt1}) is redundant. In the following we assume (\ref{rdt1}) has already been removed.

The case of (\ref{rdt2}) is a little bit involved due to the constraint (\ref{cnst1}). Let us first consider the following two statements: $\mathcal{P}$ satisfies (\ref{cnst1}); $\mathcal{P}$ satisfies (\ref{cnst1}) where $U_2$ is removed. If the latter statement is true, so is the former, but not vice versa. In the following, we will first focus on a class of $\mathcal{P}$ satisfying both.

We prove that (\ref{rdt2}) is redundant using a similar argument for (\ref{rdt1}). If (\ref{rdt2}) is violated, we have
{\small
\begin{align}
    R_3&> I(V_1;Y_1|U_1,Q) + I(V_2,U_2;Y_2|U_1) + I(X_2;Y_3|U_2)-\notag\\
    &\quad\ I(V_1;V_2|U_1). \label{violated2}
\end{align}
}If $(R_1,R_2,R_3)$ satisfies all inequalities in $\hat{\mathcal{R}}$ except (\ref{rdt2}), then $(R_1,R_2,R_3)\in\hat{\mathcal{R}}_{\mathcal{P}_{U_2=\phi}}$. Again to obtain (\ref{equi1}), we use (\ref{R2plusR3}) and (\ref{violated2}) and have
{\small
\begin{align*}
    R_2&\leq I(V_2;Y_2) - [I(V_1;Y_1|U_1)+I(V_2;Y_2|U_1)-I(V_1;V_2|U_1)]\\
    &\leq I(V_2;Y_2),
\end{align*}
}where the last inequality is due to the fact that $\mathcal{P}$ satisfies (\ref{cnst1}) where $U_2$ is removed. Similarly (\ref{equi2}) follows from (\ref{sum1}) and (\ref{violated2}) and (\ref{equi3}) from (\ref{sum2}) and (\ref{violated2}). Hence by time-sharing, (\ref{rdt2}) is redundant for this class of $\mathcal{P}$.

Next we focus on a class of $\mathcal{P}$ satisfying (\ref{cnst1}) but not (\ref{cnst1}) without $U_2$, i.e.
{\small\begin{align}
    I(V_1;Y_1|U_1)+I(V_2;Y_2|U_1)-I(V_1;V_2|U_1)<0 \label{abovecondi}.
\end{align}
}As we can see, an attempt to repeat what we have done previously fails in this case since $\mathcal{P}_{U_2=\phi}$ is not a valid joint input distribution. However, a careful examination of (\ref{abovecondi}) reveals the truth that this particular $\mathcal{P}$ is simply a bad choice for the binning coding because the penalty term $I(V_1;V_2|U_1)$ arising from having correlated inputs is so large that we might have done better provided no binning coding is used. Hence we consider $\mathcal{P}_{U_1=V_1}$, where essentially we only make use of superposition and the binning aspect is not present resulting in $I(V_1;V_2|U_1)=0$. From (\ref{abovecondi}), especially two derived conditions $I(V_1;Y_1|U_1)-I(V_1;V_2|U_1)<0$ and $I(V_2;Y_2|U_1)-I(V_1;V_2|U_1)<0$, it can be checked that $\hat{\mathcal{R}}_{\mathcal{P}} \subseteq \hat{\mathcal{R}}_{\mathcal{P}_{U_1=V_1}}$ for this class of joint distributions. Then by a similar argument for (\ref{rdt1}), it can be shown that any rates that violate (\ref{rdt2}) automatically fall into $\hat{\mathcal{R}}_{\mathcal{P}_{U_1=V_1,U_2=\phi}}$. Hence by time-sharing, (\ref{rdt2}) is redundant.
\end{IEEEproof}

\section{Proof of Corollary \ref{CoroR2}}
\label{ProofCoroR2}
\begin{IEEEproof}

Consider a region $\tilde{\mathcal{R}}$ which is the same as $\hat{\mathcal{R}}$ in Lemma \ref{R3hat}, App. \ref{ProofTheoremBinningRegion}, except that (\ref{rdt2}) is removed. Since (\ref{rdt1}) and (\ref{rdt2}) are redundant by Proposition \ref{equivalence}, App. \ref{ProofTheoremBinningRegion}, we have $\hat{\mathcal{R}}=\tilde{\mathcal{R}} = \mathcal{R}$. Now fix $Q$, evaluate $\tilde{\mathcal{R}}$ with $X_1=V_1$, $V_2=U_1$ to obtain a region specified by the same inequalities defining $\mathcal{R}_2$ plus one extra inequality,
\begin{align*}
    R_3\leq I(U_2;Y_2|U_1)+I(X_2;Y_3|U_2).
\end{align*}
Using the same argument in Proposition \ref{equivalence}, App. \ref{ProofTheoremBinningRegion}, we can show that this inequality is redundant.
\end{IEEEproof}

\section{Proof of Proposition \ref{convexification}}
\label{ProofPropositionConv}
\begin{IEEEproof}

We prove for $\mathcal{R}_1$. The case of $\mathcal{R}_2$ follows similarly.

Let $Q$ take two values 1, 2 with probability $\alpha$ and $\bar{\alpha}$, $0\leq\alpha\leq1$. Denote two sets of $(U_1^i,X_1^i,U_2^i,X_2^i,Y_1^i,Y_2^i)$ where $i=1,2$. For $Q=i$, define $U_{1,Q}=U_1^i$, $U_{2,Q}=U_2^i$, $X_1=X_1^i$, $X_2=X_2^i$, $Y_1=Y_1^i$ and $Y_2=Y_2^i$. Then we have Markov chain $(Q,U_{1,Q},U_{2,Q})\rightarrow (X_1,X_2)\rightarrow(Y_1,Y_2)$.

For the 1st inequality in $\mathcal{R}_1$,
{\small\begin{align*}
    \alpha I(U_1^1;Y_1^1) + \bar{\alpha} I(U_1^2;Y_1^2)=I(U_{1,Q};Y_2|Q)\leq I(U_{1,Q},Q;Y_2).
\end{align*}

}
For the 2nd inequality in $\mathcal{R}_1$,
{\small\begin{align*}
    \alpha I(X_2^1;Y_3^1) &+ \bar{\alpha} I(X_2^2;Y_3^2)\\
    &=I(X_2;Y_3|U_{2,Q},Q) +  I(U_{2,Q};Y_3|Q)\\
    &\leq  I(X_2;Y_3|U_{2,Q},Q) +  I(U_{2,Q},Q;Y_3)=I(X_2;Y_3)
\end{align*}

}Similarly, we can show that the convex combinations of the right-hand sides of the 3rd, 4th inequalities in $\mathcal{R}_1$ are respectively less or equal to
{\small\begin{align*}
    &I(U_{1,Q},Q;Y_1)+I(X_1;Y_2|U_{1,Q},U_{2,Q},Q)\\
    &I(U_{1,Q},Q;Y_1)+I(X_1,U_{2,Q},Q;Y_2|U_{1,Q},Q)+I(X_2;Y_2|U_{2,Q},Q)
\end{align*}

}Redefine $U_1=(U_{1,Q},Q)$ and $U_2=(U_{2,Q},Q)$. We see that the time-sharing region is always contained within $\mathcal{R}_1$ for some  $\mathcal{P}_{U_1,X_1,U_2,X_2}$.
\end{IEEEproof}

\section{Proof of Lemma \ref{DEXPR2minus}}
\label{proofDEXPR2minus}
\begin{IEEEproof}
We will use the following notations. $\Omega$ denotes the set of all DExPs of $\mathcal{R}_{2}$. For some predefined $R_1'$, denote $\mathcal{R}_{2}(R_1') =\{(R_2,R_3): (R_1',R_2,R_3) \in\mathcal{R}_{2}\}$ and the corresponding set of all DExPs $\Omega(R_1')$. Similarly, we could also define $\mathcal{R}_{2}(R_i')$ and $\Omega(R_i')$ for $i=2,3$, and $\mathcal{R}_{2}(R_k',R_l')$ and $\Omega(R_k',R_l')$ for $k,l=1,2,3$ and $k< l$.

$\mathcal{R}_{2}$ is given by a system of linear inequalities. Since DExPs are ExPs by definition, which can be found by solving the system of linear equations given by some active constraints, one approach to find $\Omega$ is to consider all possible combinations of active constraints whose corresponding system of linear equations admits a unique solution and then compare the obtained ExPs one by one. There are totally eight inequalities in $\mathcal{R}_{2}$ making this approach tedious. Fortunately, we can make use the property of DExPs to simply the procedure and make it more systematic so that we don't overlook any DExP.

Let $R_i^*$ denote the largest admissible $R_i$ in $\mathcal{R}_{2}$. Then DExPs can be sorted into four categories:

\noindent Case 1: $(R_1^*,R_2,R_3)\in\Omega$ for some $R_i\leq R_i^*$, $i=2,3$\\
Case 2: $(R_1,R_2^*,R_3)\in\Omega$ for some $R_i\leq R_i^*$, $i=1,3$\\
Case 3: $(R_1,R_2,R_3^*)\in\Omega$ for some $R_i\leq R_i^*$, $i=1,2$\\
Case 4: $(R_1,R_2,R_3)\in\Omega$ for some $R_i<R_i^*$, $i=1,2,3$

Note that Case 1, 2, 3 are not mutually exclusive. The point of a such division is, by considering Case 1, 2, 3, a higher dimensional ($n=3$) problem can be reduced to a lower one ($n=1$ or $n=2$) and for the irreducible Case 4, the additional constraints $R_i<R_i^*$ will simplify the problem. This point will be made clear as we proceed in the following.

\textit{Case 1:}

The largest admissible $R_1^*=I(X_1;Y_1|U_1)+I(U_1;Y_2|U_2)$ is obtained by setting $R_2=0$ in (\ref{largestR1}). Fixing $R_1'=R_1^*$, $R_2'=0$, the following two statements are equivalent
\begin{align*}
    (R_1',R_2',R_3)\in\Omega \Longleftrightarrow R_3\in\Omega(R_1',R_2').
\end{align*}
Since $\mathcal{R}_{2}(R_1',R_2')$ is one dimensional, we have $\Omega(R_1',R_2') = \big\{\sup_{R_3\in\mathcal{R}_{2}(R_1',R_2')} R_3 \big\}=\big\{\min\{ I(X_2;Y_3) , I(U_2;Y_2)+I(X_2;Y_3|U_2) \}\big\}$, which gives us $D$.

\textit{Case 2:}

The largest admissible $R_2^*=I(U_1;Y_2|U_2)$ is given by (\ref{largestR2}). Fixing $R_2'=R_2^*$, we have $(R_1,R_2',R_3) \in \Omega$ iff $(R_1,R_3)\in\Omega(R_2')$ and
{\small\begin{align*}
    &\mathcal{R}_{2}(R_2')=\\
    &\left\{\begin{aligned}
    &(R_1,R_3):  R_1,R_3\geq 0,\ R_1 \leq I(X_1;Y_1|U_1) \triangleq a    \\
    &R_3\leq \min\{I(X_2;Y_3),\ I(U_2;Y_2)+I(X_2;Y_3|U_2)\}  \triangleq b
    \end{aligned}\right\}
\end{align*}
}is two dimensional. It is easy to see that $\Omega(R_2')=\big\{(a,b)\big\}$ yielding $A$.

\textit{Case 3:}

The largest admissible $R_3$ is given by $R_3^*=\min\big\{I(X_2;Y_3),\ I(U_1,U_2;Y_2)+I(X_2;Y_3|U_2)\big\} $. If $R_3^*=I(X_2;Y_3)$, fixing $R_3'=R_3^*$ and we have
{\small
\begin{align*}
    &\mathcal{R}_{2}(R_3')=\\
    &\left\{\begin{aligned}
    &(R_1,R_2):  R_1,R_2\geq 0\\
    &R_2\leq \min\{I(U_1;Y_2|U_2),\ I(U_1,U_2;Y_2)-I(U_2;Y_3)\}\triangleq c  \\
    &R_1+R_2 \leq I(X_1;Y_1|U_1)+ \min\{I(U_1;Y_2|U_2),\\
    &\qquad\qquad\quad I(U_1,U_2;Y_2)-I(U_2;Y_3)\} \triangleq d  \\
    \end{aligned}\right\}.
\end{align*}
}Note that $I(U_1,U_2;Y_2)-I(U_2;Y_3)\geq 0$ in this case. It is easy to see $\Omega(R_3')=\big\{(d-c,c),(d,0)\big\}$, resulting in two DExPs:
{\small\begin{align*}
    B'&=\big( I(X_1;Y_1|U_1),\ \min\{I(U_1;Y_2|U_2),\ I(U_1,U_2;Y_2)-\\
    &\quad \ \ I(U_2;Y_3)\},\ I(X_2;Y_3)  \big)\\
    C'&=\big( I(X_1;Y_1|U_1)+ \min\{I(U_1;Y_2|U_2),\ I(U_1,U_2;Y_2)-\\
    &\quad \ \ I(U_2;Y_3)\},\ 0, \  \ I(X_2;Y_3)  \big)
\end{align*}
}If $R_3^*=I(U_1,U_2;Y_2)+I(X_2;Y_3|U_2)$, which is given by (\ref{largestR3}) by setting $R_2=0$, fixing $R_2'=0$, $R_3'=R_3^*$ and we obtain $\mathcal{R}_{2}(R_2',R_3')=\big\{ R_1: 0\leq R_1\leq  I(X_1;Y_1|U_1)  \big\}$ and $\Omega(R_2',R_3')=\big\{I(X_1;Y_1|U_1)\big\}$. Hence we obtain one DExP
{\small\begin{align*}
    E'=\big( I(X_1;Y_1|U_1),\ 0,\ I(U_1,U_2;Y_2)+I(X_2;Y_3|U_2)  \big).
\end{align*}
}Combing the two cases, we rewrite $B'$ and $E'$ collectively as $B$ and $C'$ and $E'$ collectively as $C$.

\textit{Case 4:}

Under the condition $R_i<R_i^*$, $i=1,2,3$, $\mathcal{R}_{2}$ is given by
{\small\begin{align}
    R_1&<I(X_1;Y_1|U_1)+I(U_1;Y_2|U_2)  \label{vioR1}\\
    R_2&< I(U_1;Y_2|U_2)\label{vioR2}\\
    R_3&< \min\{ I(X_2;Y_3),\ I(U_1,U_2;Y_2)+\notag\\
    &\quad\ I(X_2;Y_3|U_2) \} \label{vioR3}\\
    R_1+R_2&\leq I(X_1;Y_1|U_1)+I(U_1;Y_2|U_2)\label{constraint1}\\
    R_2+R_3&\leq I(U_1,U_2;Y_2)+I(X_2;Y_3|U_2)\label{constraint2}\\
    R_1+R_2+R_3&\leq I(X_1;Y_1|U_1)+I(U_1,U_2;Y_2)+\notag\\
    &\quad\ I(X_2;Y_3|U_2)\label{constraint3}\\
    R_1,R_2,R_3&\geq 0.  \label{constraint4}
\end{align}
}As mentioned before, DExPs are the solutions of systems of linear equations given by some active constraints. Hence we first consider all possible combinations of active constraints defining dominant faces, i.e. (\ref{constraint1})-(\ref{constraint3}) and then add additional active constraints from (\ref{constraint4}) as needed to ensure the resulting system has a unique solution.

If (\ref{constraint1}), (\ref{constraint2}), (\ref{constraint3}) are all active, from (\ref{constraint2}), (\ref{constraint3}) we get $R_1=I(X_1;Y_1|U_1)$ and further with (\ref{constraint1}), we get $R_2=I(U_1;Y_2|U_2)$, which violates (\ref{vioR2}). 
If only (\ref{constraint1}), (\ref{constraint2}) are active, since the corresponding system of linear equations does not have a unique solution (more variables than equations), we choose one additional active constraint from (\ref{constraint4}). However, the obtained solution violates either (\ref{vioR1}), (\ref{vioR2}) or (\ref{vioR3}). 
We can proceed similarly for the remaining six possible combinations 
and none of them yields a valid DExP. Overall we conclude that there is no DExP in Case 4.
\end{IEEEproof}

\section{Proof of Lemma \ref{DEXPRpprimeminus}}
\label{proofDEXPRpprimeminus}
\begin{IEEEproof}
We use the same notations from the proof of Lemma \ref{DEXPR2minus}, which are now defined over $\mathcal{R}_{(2)}$ instead of $\mathcal{R}_{2}$. As we can see, $R_1$ is disassociated with $R_2$, $R_3$. Hence the DExPs of $\mathcal{R}_{(2)}$ is of the form $(R_1^*,R_2,R_3)$, where $R_1^*=I(X_1;Y_1|U_1)$. Fixing $R_1'=R_1^*$, 
we have
{\small\begin{align}
    \mathcal{R}_{(2)}(R_1')=
    &\left\{\begin{aligned}
    &(R_2,R_3): R_2,R_3\geq 0 \notag\\
    &R_2\leq I(U_1;Y_2|U_2)\triangleq a,\ R_3\leq I(X_2;Y_3) \triangleq b \notag\\
    &R_2+R_3\leq I(U_1,U_2;Y_2)+I(X_2;Y_3|U_2) \triangleq c \notag\\
    \end{aligned}\right\}.
\end{align}

}
If $c\geq b$, i.e. $I(U_1,U_2;Y_2)\geq I(U_2;Y_3)$, $\mathcal{R}_{(2)}(R_1')$ is depicted in Fig. \ref{Region}.(a) and $\Omega(R_1')=\{T_1,T_2\}=\{(a,\min\{b,c-a\}),(\min\{a,c-b\},b)\}$, yielding two DExPs, $A$ and
{\small\begin{align*}
    B'&=\Big( I(X_1;Y_1|U_1),\ \min\{I(U_1;Y_2|U_2),\ I(U_1,U_2;Y_2)-\\
    &\qquad I(U_2;Y_3) \} ,\  I(X_2;Y_3) \Big).
\end{align*}

}
If $c< b$, i.e. $I(U_1,U_2;Y_2)< I(U_2;Y_3)$, $\mathcal{R}_{(2)}(R_1')$ is depicted in Fig. \ref{Region}.(b) and $\Omega(R_1')=\{T_1,T_3\}=\{(a,\min\{b,c-a\}),(0,c)\}$, yielding one more DExP of $\mathcal{R}_{(2)}$
{\small\begin{align*}
    C'&=\Big( I(X_1;Y_1|U_1),\ 0 ,\ I(U_1,U_2;Y_2)+I(X_2;Y_3|U_2) \Big).
\end{align*}

}
Note that $B'$ and $C'$ can be rewritten collectively as $B$.
\begin{figure}
  \centering
  \subfloat[]{\includegraphics[width=0.15\textwidth]{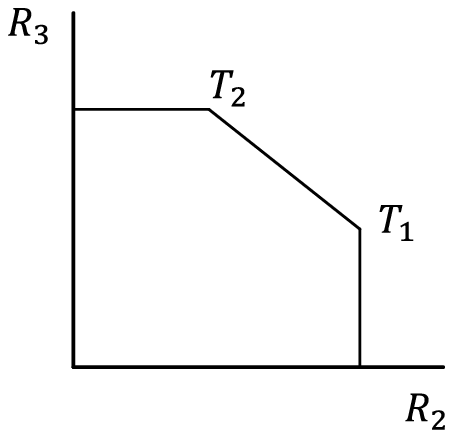}}
  \subfloat[]{\includegraphics[width=0.15\textwidth]{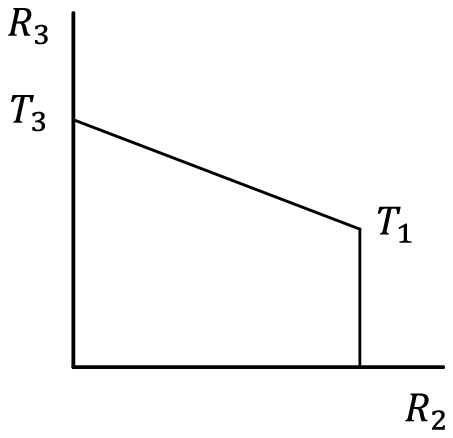}}
  \caption{}
  \label{Region}
\end{figure}
\end{IEEEproof}

\section{Proof of Lemma \ref{DEXPR1minus}}
\label{proofDEXPR1minus}
\begin{IEEEproof}
Since the argument is similar to that of Lemma \ref{DEXPR2minus}, we use the same notations, which are now defined over $\mathcal{R}_{1}$ instead of $\mathcal{R}_{2}$. Again the DExPs can be sorted into four categories. We next discuss case by case.

\textit{Case 1:}

The largest admissible $R_1^*=I(U_1;Y_1)$. Fixing $R_1'=R_1^*$, we have $(R_1',R_2,R_3) \in \Omega$ iff $(R_2,R_3) \in \Omega(R_1')$ and
{\small\begin{align*}
    \mathcal{R}_{1}(R_1')=
    &\left\{\begin{aligned}
    &(R_2,R_3):  R_2,R_3\geq 0\\
    &R_2\leq I(X_1;Y_2|U_1,U_2),\ R_3\leq I(X_2;Y_3)  \\
    &R_2+R_3\leq I(X_1,U_2;Y_2|U_1)+I(X_2;Y_3|U_2)
    \end{aligned}\right\}.
\end{align*}
}Similar to the proof of Lemma \ref{DEXPRpprimeminus}, there are two DExPs: $E$, $F$.

\textit{Case 2:}

The largest admissible $R_2^*=I(U_1;Y_1)+I(X_1;Y_2|U_1,U_2)$ is obtained by setting $R_1=0$ in (\ref{largestR22}). Fixing $R_1'=0$ and $R_2'=R_2^*$, we have $\mathcal{R}_{1}(R_1',R_2')=\big\{R_3:0\leq R_3\leq \min\{ I(X_2;Y_3) ,\ I(U_2;Y_2|U_1)+I(X_2;Y_3|U_2) \}\big\}$, resulting in point $G$.

\textit{Case 3:}

The largest admissible $R_3^*=\min\{ I(X_2;Y_3),\ I(U_1;Y_1)+I(X_1,U_2;Y_2|U_1)+I(X_2;Y_3|U_2)\}$.

{\em 1.} If $I(U_2;Y_3)\leq I(U_1;Y_1)+I(X_1,U_2;Y_2|U_1)$, $R_3^*=I(X_2;Y_3)$. Fixing $R_3'=R_3^*$, we have
\begin{align*}
    &\mathcal{R}_{1}(R_3')=\\
    &\left\{\begin{aligned} &(R_1,R_2):R_1,R_2\geq 0,\  R_1\leq I(U_1;Y_1) \triangleq a\\
    &R_1+R_2 \leq I(U_1;Y_1)+ I(X_1;Y_2|U_1,U_2)+ \\
    &\qquad\qquad\quad \min\{0,\ I(U_2;Y_2|U_1)-I(U_2;Y_3)\} \triangleq b\\
    \end{aligned}\right\}.
\end{align*}
Note that $b\geq 0$ in this case. This case is similar to that in Lemma \ref{DEXPRpprimeminus} and we can show
\begin{enumerate}
\item If $I(U_2;Y_3)\leq I(X_1,U_2;Y_2|U_1)$, $\Omega(R_3')=\{(0,b),(a,b-a)\}$
\item If $I(X_1,U_2;Y_2|U_1)<I(U_2;Y_3)\leq I(U_1;Y_1)+I(X_1,U_2;Y_2|U_1)$, $\Omega(R_3')=\{(0,b),(b,0)\}$
\end{enumerate}
Finally, we collectively write the obtained DExPs as
{\small
\begin{align*}
    H =\Big(0,\ b,\ R_3^* \Big),\ I = \Big(\min\{a,b\},\ [b-a]^+,\ R_3^* \Big).
\end{align*}

}
{\em 2.} If $I(U_2;Y_3)> I(U_1;Y_1)+I(X_1,U_2;Y_2|U_1)$, $R_3^*=I(U_1;Y_1)+I(X_1,U_2;Y_2|U_1)+I(X_2;Y_3|U_2)$, which is obtained by setting $R_1=R_2=0$ in (\ref{largestR33}). In this case we find one DExP $J$.

\textit{Case 4:} Similar to that for Lemma \ref{DEXPR2minus}, it can be shown that there is no DExP in this case.
\end{IEEEproof}

\section{Proof of Theorem \ref{capastrong}}
\label{proofcapastrong}
To prove the converse, we use the technique proposed in \cite{Nair}. Specifically, we need the following lemma that can be easily proved using the same arguments for \cite[Lemma 1]{Nair}.

\begin{lemma}
\label{lemmalessnoisy}
In a DM-BIC with $Y_1\prec_o Y_2$, if $W\rightarrow(X_1^n,X_2^n)\rightarrow(Y_1^n,Y_2^n)$ form a Markov chain, then the following holds:
\begin{align*}
    I(Y_2^{i-1};Y_{2,i}|W)\geq I(Y_1^{i-1};Y_{2,i}|W),\ 1\leq i \leq n.
\end{align*}
\end{lemma}

\begin{IEEEproof}[Proof of Theorem \ref{capastrong}]

The achievable schemes are given by the coding schemes for $\mathcal{R}_1$ in Corollary \ref{CoroR1} and $\mathcal{R}_{(1)}$ in Theorem \ref{Rprime} respectively, all with $U_2=X_2$.

For the converse, we define $U_i=(W_1,Y_1^{i-1})$. For some $\epsilon_n$ such that $\lim_{n\rightarrow\infty}\epsilon_n=0$, by Fano's inequality, we have
{\small\begin{align*}
    n(R_1-\epsilon_n)&\leq I(W_1;Y_1^n)\\
    &=\sum_{i=1}^n I(W_1;Y_{1,i}|Y_1^{i-1}) \leq \sum_{i=1}^n I(U_i;Y_{1,i}).
\end{align*}

}To bound $R_2$, we proceed as the follows
{\small\begin{align*}
    n(R_2-\epsilon_n)&\leq I(W_2;Y_2^n|W_1,X_2^n)\\
    &\leq \sum_{i=1}^n I(X_{1,i};Y_{2,i}|W_1,X_{2,i},Y_2^{i-1})\\
    &=\sum_{i=1}^n I(X_{1,i};Y_{2,i}|W_1,X_{2,i})-I(Y_2^{i-1};Y_{2,i}|W_1,X_{2,i})\\
    &\overset{(a)}{\leq} \sum_{i=1}^n I(X_{1,i};Y_{2,i}|W_1,X_{2,i})-I(Y_1^{i-1};Y_{2,i}|W_1,X_{2,i})\\
    &=\sum_{i=1}^n I(X_{1,i};Y_{2,i}|U_i,X_{2,i}),
\end{align*}
}where $(a)$ follows from Lemma \ref{lemmalessnoisy}.

Now we consider upper-bound for $R_2+R_3$. The strong condition implies $I(X_2^n;Y_2^n|X_1^n)\geq I(X_2^n;Y_3^n)$, \cite[Lemma]{Costa}. Proceeding,
{\small\begin{align*}
    n(R_2+R_3-\epsilon_n)&\leq I(W_2;Y_2^n)+I(W_3;Y_3^n)\\
    &\overset{(b)}{\leq} I(X_1^n;Y_2^n|W_1) + I(X_2^n;Y_2^n|X_1^n)\\
    &=\sum_{i=1}^n I(X_{1,i},X_{2,i};Y_{2,i}|W_1,Y_2^{i-1})\\
    &=\sum_{i=1}^n I(X_{1,i},X_{2,i};Y_{2,i}|W_1)- I(Y_2^{i-1};Y_{2,i}|W_1)\\
    &\overset{(c)}{\leq} \sum_{i=1}^n I(X_{1,i},X_{2,i};Y_{2,i}|W_1)- I(Y_1^{i-1};Y_{2,i}|W_1)\\
    &=\sum_{i=1}^n I(X_{1,i},X_{2,i};Y_{2,i}|U_{i}),
\end{align*}
}where $(b)$ is due to the strong condition and $(c)$ is due to Lemma \ref{lemmalessnoisy}.

Finally, we have $n(R_3-\epsilon_n)\leq \sum_{i=1}^n I(X_{2,i};Y_{3,i})$. The proof is complete by redefining $U=(U_Q,Q)$, $X_{j,i}=X_{j}$ for $j=1,2$, and $Y_{l,i}=Y_{l}$, for $l=1,2,3$, where $Q$ is a uniformly distributed r.v. on $(1,...,n)$.
\end{IEEEproof}



\begin{thebibliography}{1}

\bibitem{Marton}
K. Marton, ``A coding theorem for the discrete memoryless broadcast channel,'' \textit{IEEE Trans. Inf. Theory}, vol. 25, no. 3, pp. 306-11, 1979.

\bibitem{Han}
T. S. Han and K. Kobayashi, ``A new achievable rate region for the interference channel,'' \textit{IEEE Trans. Inf. Theory}, vol. 27, no. 1, pp. 49-60, Jan. 1981.

\bibitem{Cover BC}
T. M. Cover, ``Broadcast channels,'' \textit{IEEE Trans. Inf. Theory}, vol. 18, no. 1, pp. 2-14, Jan. 1972.

\bibitem{Korner}
J. Korner and K. Marton, ``Comparison of two noisy channels,'' in \textit{Topics in Information Theory (Second Colloq., Keszthely, 1975)} Amsterdam: North-Holland, 1977, pp. 411-23.

\bibitem{El Gamal Costa}
A. El Gamal and M. Costa, ``The capacity region of a class of deterministic interference channels,'' \textit{IEEE Trans. Inf. Theory}, vol. 28, no. 2, pp. 343-46, 1982.

\bibitem{Etkin}
R. Etkin, D. Tse, and H. Wang, ``Gaussian interference channel capacity to within one bit,'' \textit{IEEE Trans. Inf. Theory}, vol. 54, no. 12, pp. 5534-562, Dec. 2008.

\bibitem{Andrews}
J. G. Andrews, H. Claussen, M. Dohler, S. Rangan, M. C. Reed, ``Femtocells: past, present, and future,'' to appear \textit{IEEE JSAC on Femtocellular Networks}, Apr 2012.

\bibitem{Shang I}
X. Shang and H. V. Poor, ``On the capacity of type I broadcast-Z-interference channels," \textit{IEEE Trans. Inf. Theory}, Special Issue on Interference Networks, vol. 57, no. 5, pp. 2648-66, May 2011.

\bibitem{Shang II}
X. Shang and H. V. Poor, "On the capacity of Gaussian broadcast channels that receive interference," in \textit{Proc. 45th Conference on Information Sciences and Systems}, Baltimore, MD, pp. 1-5, March 2011.

\bibitem{Cover}
T. Cover and J. Thomas, \textit{Elements of Information Theory 2nd Edt}, New York: Wiley, 2006.

\bibitem{El Gamal}
A. El Gamal and Y. Kim, \textit{Network Information Theory}, Cambridge University Press, 2011.


\bibitem{CMG}
H. Chong,  M. Motani, H. Garg and H. El Gamal, ``On the Han-Kobayashi region for the interference channel,'' \textit{IEEE Trans. Inf. Theory}, vol. 54 no.7, p. 3188-95, 2008.

\bibitem{Costa}
M. Costa and A. El Gamal, ``The capacity region of the discrete memoryless interference channel with strong interference,'' \textit{IEEE Trans. Inf. Theory}, vol. 33, no. 5, pp. 710-11, September 1987.

\bibitem{Nair}
C. Nair and Z. V. Wang, ``The capacity region of the three receiver less noisy broadcast channel,'' \textit{IEEE Trans. Inf. Theory}, vol. 54 no.7, pp. 4058-62, 2011.
\end{thebibliography}
\end{document}